\documentclass[twocolumn,prb,showpacs,multicol,amsmath,amssymb]{revtex4}
\usepackage[dvips]{graphicx}
\usepackage{graphicx}
\usepackage{dcolumn}
\usepackage{bm}
\usepackage{graphics}
\usepackage{epsfig,color}
\usepackage[normalem]{ulem} 
\usepackage{soul}

\newcommand{\be}{\begin{equation}}
\newcommand{\ee}{\end{equation}}
\newcommand{\bea}{\begin{eqnarray}}
\newcommand{\eea}{\end{eqnarray}}

\begin{document}
\title{Ineffectiveness of the Dzyaloshinskii-Moriya interaction in the dynamical quantum phase transition in the ITF model}
\author{Hadi Cheraghi$^{1}$}
\email[]{h.cheraghi1986@gmail.com}
\author{Saeed Mahdavifar$^{2}$}
\email[]{smahdavifar@gmail.com}
\affiliation{$^{1}$ Department of Physics, Semnan University, 35195-363, Semnan,
Iran}
\affiliation{$^{2}$ Department of Physics, University of Guilan, 41335-1914, Rasht, Iran}

\date{\today}
\begin{abstract}
Quantum phase transition occurs at a quantum critical value of a control parameter such as the magnetic field in the Ising model in a transverse magnetic field (ITF). Recently, it is shown that ramping across the quantum critical point generates  non-analytic behaviors in the time evolution of a closed quantum system in the thermodynamic limit at zero temperature. The mentioned phenomenon is called the dynamical quantum phase transition (DQPT). Here, we consider the one-dimensional (1D) ITF model with added the Dzyaloshinskii-Moriya  interaction (DMI). Using the fermionization technique, the Hamiltonian is exactly diagonalized. Although the DM interaction induces chiral phase in the ground state phase diagram of the model, the study of the rate function of the return probability has proven that the DMI does not affect in the DQPT. We conclude accordingly that the ramping across the quantum critical point is not  a necessary and sufficient condition for DQPT.

\end{abstract}
\maketitle

\textbf{Introduction-}
In quantum many body systems, recognition of dynamical effects created by different interactions between particles is very exciting. In recent years, people have focused on some low-dimensional quantum magnets where include a variety of interactions.   The creation of many techniques for building different optical lattices has helped to know new Hamiltonians which have caused the expansion of research in the field of quantum magnets. One-dimensional spin-1/2 systems are a  practical category for studying specially Ising model in a transverse field (ITF) [{\color{blue}\onlinecite{{Blatt},{Friedenauer}}}] and XXZ model [{\color{blue}\onlinecite{{Zeitschrift},{Dmitriev1},{Toskovic}}}] because, despite their simple and exactly solvable Hamiltonian, low energy behavior of many systems can be realized through them. Such interactions that break the rotation symmetry, reduce the quantum fluctuations and can tend to induce ground state magnetic phases.

The antisymmetric spin exchange interaction, known as Dzyaloshinskii-Moriya  interaction (DMI) is responsible for the presence of weak ferromagnetism in a variety of antiferromagnetic models. The competition between the Heisenberg exchange interaction and the DMI leads to the formation of exotic structures such as chiral domain walls [{\color{blue}\onlinecite{{Miron},{Thiaville}}}], helices [{\color{blue}\onlinecite{{Grigoriev1},{Grigoriev2}}}], and skyrmions [{\color{blue}\onlinecite{{Romming},{Pollard},{Fert}}}] that have attracted the interest of both theorists and experimentalists in condensed matter physics [{\color{blue}\onlinecite{{Oshikawa},{Tsukada},{Kohgi},{Herak},{Gitgeatpong},{Guoqiang}}}]. Dzyaloshinskii [{\color{blue}\onlinecite{Dzyaloshinskii}}] has shown that in crystal with the lack of structural inversion symmetry, the usual isotropic exchange interaction $(J{\overrightarrow S _i}.{\overrightarrow S _j})$ is not only magnetic interaction but also antisymmetric exchange interaction $(\overrightarrow D {\overrightarrow {.S} _i} \times {\overrightarrow S _j})$  . Later, Moriya has shown that inclusion of spin orbit coupling on magnetic ions in the 1st and 2nd order leads to antisymmetric and anisotropic exchange, respectively [{\color{blue}\onlinecite{Moriya60}}].

The DMI is important for stabilizing non-collinear magnetic structures in ferromagnets [{\color{blue}\onlinecite{{Zhang},{Rigol2}}}]. Historically, first, they have been  considered in the context of $ZnCu3(OH)6Cl3$ [{\color{blue}\onlinecite{Helton}}] to explain the enhancement of the spin susceptibility at low temperatures. Theoretically, the study of induced effects of the DMI on the ground state and the finite temperature behavior of the low-dimensional magnets has attracted much interest in recent years [{\color{blue}\onlinecite{Oshikawa, Jafari08, Kargarian09, Soltani10, Li09, Mahdavifar10, Karimi11, Soltani12, Vahedi12,  Hasanzadeh14, Mahdavifar15, Chan17}}].

It is more than one decade that we have seen a tremendous  interest in the physics for considering and understanding of  the non-equilibrium dynamics in quantum many-body systems. This  issue has been developed remarkably after some  experimental
progress in cold atom systems as a fundamental concept known as $ quantum~quench$ [{\color{blue}\onlinecite{{Greiner},{Chen},{Kollath}}}]. In sudden quantum quench, the system is initially prepared typically in   an equilibrium  ground state $\left| {{\Psi _i}} \right\rangle $ of an initial Hamiltonian $H_i$. At time t = 0, suddenly we switch a parameter control of the system from  its initial value to final value so that the final Hamiltonian will be $H_f$. After that, the system evolves with passing time [{\color{blue}\onlinecite{{Calabrese1},{Fagotti1},{Fagotti2}}}]. The Loschmidt echo (LE) is a good candidate for studying the non-equilibrium time evolution of quantum systems after doing quench [{\color{blue}\onlinecite{{Quan},{Jafari1}}}] and defines as
\begin{equation}
G(t) =  {\left\langle {{\Psi _i}} \right|{e^{ - iH_ft}}\left| {{\Psi _i}} \right\rangle }.
\end{equation}
The LE is a measure of the stability of the  time-reversal of a system. Therefore, it can be used to quantify the decoherence effects in quantum systems [{\color{blue}\onlinecite{{Cucchietti},{Gorin}}}]. Hence, prior works usually were focused on investigating this certain scenario. In 2013,  Heyl et al [{\color{blue}\onlinecite{Heyl1}}] have discovered the formal similarity of the canonical partition function of an equilibrium system, $Z(\beta ) = tr\left( {{e^{ - \beta H}}} \right)$, and the LE, called $dynamical~quantum~phase~transition(DQPT)$, that denotes  non-analytic behaviors of the system in critical points in the real time evolution. In the thermodynamic limit, they derived a similarity between the free energy density of system, $f(z) =  - \mathop {\lim }\limits_{N \to \infty } \frac{1}{N}\log \left( {Z(z)} \right)$, with $z \in \mathbb{C}$ in complex temperature plane and  the rate function of the return probability which is given by
\begin{equation}
l(t) = f( it) +f(-it)=  - \mathop {\lim }\limits_{N \to \infty } \frac{1}{N}\log   \left|{G(t)} \right|^2.
\end{equation}
The DQPTs can be recognized by analyzing the dynamics of the rate function of the return probability where this leads to a non-analytic temporal behavior when quenches across the quantum critical points. Furthermore, DQPTs occur in the critical times where the Fisher zeroes exist. It should be noted that DQPTs have been observed and verified in different experiments recently [{\color{blue}\onlinecite{{Heyl2},{Zhang},{Fläschner},{Martinez},{Smith}}}].

It has been shown that the mentioned approach works exactly for Hamiltonians which will be diagonalized by applying Bogoliobov transformations [{\color{blue}\onlinecite{{Heyl1},{Schmitt1},{Kehrein1}}}]. In this work, we take a first contradictory step towards this direction.
In this way, we investigate DQPTs in the spin-1/2 ITF model in presence of DMI. On the other hand, we invoke that existence of the DMI does not have any effects in DQPTs  and the system behaves exactly the same as a ITF model. It means,  Heyl's approach can not reveal effect of DMI in this model while Hamiltonian is diagonalized with the use of  analytical spinless fermion approach by applying Bogoliobov transformations.

\textbf{The model and DQPT-}
We consider the well known spin-1/2 Ising chain in a transverse magnetic field with added transverse DMI, which
is equivalent to a DM vector perpendicular to the Ising interaction axis. The Hamiltonian of the model is written as
\begin{eqnarray}
H =  {H_{DMI}} +{H_{ITF}},
\end{eqnarray}
 where
\begin{eqnarray}
H_{DMI}=\sum_{n}\mathbf{D}\cdot \left(\mathbf{S}_{n}\times \mathbf{S}_{n+1}\right),\nonumber\\
H_{ITF}=J\sum_{n}S^{x}_{n}S^{x}_{n+1} - h\sum_{n}S^{z}_{n}.
\end{eqnarray}

$\mathbf{S}_{n}$ denotes the spin-1/2 operator on the $n$-th site, $h$ is the transverse magnetic field and $J>0$ denotes antiferromagnetic coupling constant.  By considering uniform DM vector as $\mathbf{D}=D\hat{z}$, and implementing the following Jordan-Wigner transformations [{\color{blue}\onlinecite{Jordan}}]

\begin{eqnarray}
S^{+}_{n}&=&a_{n}^{\dag}e^{i\pi\sum^{n-1}_{m=1}a^{\dag}_{m}a_{m}},\nonumber\\
S^{-}_{n}&=&e^{-i\pi\sum^{n-1}_{m=1}a^{\dag}_{m}a_{m}}a_{n},\nonumber\\
S^{z}_{n}&=&a_{n}^{\dag}a_{n}-\frac{1}{2},
\label{eq5}
\end{eqnarray}
Hamiltonian takes the fermionic form as
\begin{eqnarray}
H&=&\frac{i D}{2}\sum_{n}\left(a^{\dag}_{n}a_{n+1}-a^{\dag}_{n+1}a_{n}\right)\nonumber\\
     &+& \frac{J}{4} \sum_{n} \left(a^{\dag}_{n}a^{\dag}_{n+1}-
a_{n}a_{n+1}+a^{\dag}_{n}a_{n+1}+a^{\dag}_{n+1}a_{n}\right)\nonumber\\
      &-&h\sum_{n}a^{\dag}_{n}a_{n}.
\end{eqnarray}
By performing a Fourier transformation into the momentum space as $a_{n} = \frac{1}{\sqrt{N}} \sum_{k} e^{ikn} a_{k}$, the Hamiltonian is transformed into the momentum space as
\begin{eqnarray}
H&=&\sum_{k}[D \sin(k)+\frac{J}{2}-h] a^{\dag}_{k}a_{k}\nonumber\\
     &+& \frac{iJ}{4} \sum_{k} \sin(k)\left(a^{\dag}_{-k}a^{\dag}_{k}+
a_{-k}a_{k}\right).
\end{eqnarray}
Finally,  using the Bogoliubov  transformation [{\color{blue}\onlinecite{Lieb}}]
\begin{eqnarray}
a_{k}=\cos(\theta_{k}) \beta_k+i \sin(\theta_{k})\beta^{\dag}_{-k},\nonumber\\
\end{eqnarray}
the diagonalized Hamiltonian is obtained as
\begin{eqnarray}
H = \sum\limits_{k > 0} {\left[ \varepsilon_k \beta_k^{\dag} {\beta_k} + \varepsilon_{-k} \beta _{ - k}^\dag {\beta _{ - k}} \right]},
\end{eqnarray}
where  the energy spectrum is
\begin{eqnarray}\label{eq8}
\varepsilon_k &=& \lambda (k) + B(k), \nonumber \\
A(k)&=&\frac{J}{2} \cos(k)-h, \nonumber \\
B(k)&=&-D \sin(k), \nonumber \\
C(k)&=&\frac{J}{2} \sin(k),
\end{eqnarray}
and  $\tan (2{\theta _k}) =  - \frac{{C(k)}}{{A(k)}}$. $\lambda (k) = \sqrt {{A^2}(k) + {C^2}(k)} $ corresponds to the part of the energy spectrum of system that belongs to ITF.  From Eq.~(\ref{eq8}), it is clear that in presence of the DMI, $\varepsilon_k \ne \varepsilon_{- k}$.

In absence of DMI, the model transfers to ITF model  which has a quantum phase transition at $h_c=\frac{J}{2}$ so that separates a ferromagnetic (FM) phase for $h<\frac{J}{2}$ from a paramagnetic (PM) phase for $h>\frac{J}{2}$. 
Presence of DMI by breaking symmetry of Hamiltonian changes the critical points and under given condition can induce chiral phase in the system.
The system is at its criticality when the energy gap vanishes. Using the equation, $\frac{d \varepsilon_{k}}{d k}\mid_{k_0}=0$, the energy gap wave vector $k_0$ is obtained as
\begin{eqnarray}
k_0(D=0)=0, \nonumber \\
k_0(h=0)=\frac{\pi}{2}.
\end{eqnarray}
We found that the gap of the spectrum vanishes at the critical values
\begin{eqnarray}
h_c(D=0)=\frac{J}{2}, \nonumber \\
D_c(h=0)=\frac{J}{2}.
\end{eqnarray}
The ground state of the system corresponds to the configuration where all the states with $\varepsilon_k \leq 0$ are filled and $\varepsilon _k > 0$ are empty. In this model it happens when  system is in chiral phase. Fermi points are given as
\begin{eqnarray}
\pm k_F(D=0)&=&\arccos (\frac{J}{4 h}+\frac{h}{J}), \nonumber \\
k_F^{-}(h=0)&=& \arcsin (\frac{J}{2 D}), \nonumber \\
k_F^{+}(h=0)&=& \pi-\arcsin (\frac{J}{2 D}).
\end{eqnarray}
In fact, the ground state corresponds to the configuration when all states with $|k|\leq k_F$ are filled. In the absence of the DMI, it is obvious that one Fermi point exists at zero momentum for the special value of the transverse field $h=\frac{J}{2}$. When the transverse magnetic field is absent, Fermi points can be found in the region $D \geq \frac{J}{2}$. In the presence of the transverse magnetic field and the DMI, Fermi points are obtained as
\begin{eqnarray}
k_F^{-}=\arccos (\frac{J h}{2 D^{2}}+ \sqrt{(1-\frac{J^2}{4 D^2})(1-\frac{h^2}{D^2})}, \nonumber \\
k_F^{+}=\arccos (\frac{J h}{2 D^{2}}- \sqrt{(1-\frac{J^2}{4 D^2})(1-\frac{h^2}{D^2})}~. \nonumber \\
\end{eqnarray}
It should be noted that to emerge of these two Fermi points both the conditions $D\geq h$ and $D\geq \frac{J}{2}$ must be satisfied. In order to probe the role of DMI  in dictating its influence in DQPTs of the system, first, we prepare the system in the equilibrium ground state  $\left| {{\Psi _i}(D_i,J_i,h_i)} \right\rangle $  of an initial Hamiltonian ${H_i} = H(D_i,J_i,h_i)$.  At $t=0$, we suddenly change the control parameters as $(D_i,J_i,h_i) \to (D_f,J_f,h_f)$ so that the final Hamiltonian will be ${H_f} = H(D_f,J_f,h_f)$. Now, we let the system  evolves with time as
 \begin{eqnarray}
\left| {{\Psi }(D_f,J_f,h_f)(t)} \right\rangle  = {e^{ - iH_ft}}\left| {{\Psi _i}(D_i,J_i,h_i)} \right\rangle
\end{eqnarray}
Let ${\left| 0 \right\rangle _{{\beta _k}}}$ and ${\left| 0 \right\rangle _{{\eta _k}}}$ denote the vacuum ground state  of the system before and after quench, respectively. Then, the diagonalized Hamiltonians  before and after quench can be expressed as
\begin{eqnarray}
H_i = \sum\limits_{k > 0} {\left[ \varepsilon_k(D_i,J_i,h_i)\beta _k^\dag {\beta _k} + \varepsilon_{-k}(D_i,J_i,h_i)\beta _{ - k}^\dag {\beta _{ - k}} \right]}. \nonumber
\end{eqnarray}
\begin{eqnarray}
H_f = \sum\limits_{k > 0} {\left[ \varepsilon_k(D_f,J_f,h_f)\eta _k^\dag {\eta _k} + \varepsilon_{-k}(D_f,J_f,h_f)\eta _{ - k}^\dag {\eta _{ - k}} \right]}.\nonumber \\
\end{eqnarray}

The ${\left| 0 \right\rangle _{{\beta _k}}}$ is related to the ${\left| 0 \right\rangle _{{\eta _k}}}$ through [{\color{blue}\onlinecite{{Silva1},{Fagotti1}}}]
\begin{equation}
{\left| 0 \right\rangle _{{\beta _k}}} = {\kappa ^{ - 1}}{e^{ - i\sum\limits_{k > 0} {\tan ({\Phi _k})\eta _k^\dag \eta _{ - k}^\dag } }}{\left| 0 \right\rangle _{{\eta _k}}},
\end{equation}
that ${\kappa  ^2} = \prod\limits_{k > 0} {\left( {1 + {{\tan }^2}({\Phi _k})} \right)} $ and ${\Phi _k} = {\theta _k}(D_f,J_f,h_f) - {\theta _k}(D_i,J_i,h_i)$ is the difference between the Bogoliubov angles diagonalizing the pre-quench and post-quench the Hamiltonian. It should be noticed that  existence of the DMI does not affect in the Bogoliubov angles. However, it changes the energy spectrum of the system. It is then straightforward to show that the LE is given by
\begin{eqnarray}
&G&(t)= \prod\limits_{k > 0} [ {\cos }^2(\Phi _k) \nonumber \\
 &+&{\sin }^2(\Phi _k)e^{ - it[\varepsilon _k (D_f,J_f,h_f)+\varepsilon _{-k}(D_f,J_f,h_f)]}].
\end{eqnarray}

Hence, it takes
\begin{equation}
G(t) = \prod\limits_{k > 0} {\left( {{{\cos }^2}({\Phi _k}) + {{\sin }^2}({\Phi _k}){e^{ - 2it{\lambda _k}({J_f,h_f})}}} \right)},
\end{equation}
and the rate function  of the return probability as
\begin{equation}\
l(t) =  - \frac{2}{N}\sum\limits_{k > 0} {\log } \left| {{{\cos }^2}({\Phi _k}) + {{\sin }^2}({\Phi _k}){e^{ - 2it{\lambda _k}(J_f,h_f)}}} \right|.
\end{equation}

All of them are the same as the ITF model where  the periodically critical times are $t_n^* = \frac{\pi }{{{\lambda _{{k^*}}}(J_f,h_f)}}(n + \frac{1}{2})$,$~$ $n = 0,1,2,...$, which $k^*$ is determinated by  $\cos ({k^*}) = \frac{{{J_i}{J_f} + 4{h_i}{h_f}}}{{2({J_i}{h_f} + {J_f}{h_i})}}$ [{\color{blue}\onlinecite{{Heyl1},{Kehrein1},{Zvyagin}}}].


\textbf{Summary-}
The rate function of the return probability as a good candidate for considering DQPT in quantum systems has attracted the attention of the condensed matter physicists because of its ability to detect quantum critical points without  a priori knowledge of the order parameter of the system, which is the usual way of probing a quantum phase transition. On the other hand, in contrast, some results have showed that existence or non-existence of the Fisher zeroes is not good criterion for recognizaton of DQPTs in a system. For example,
using a numerical density matrix renormalization group algorithm, it is showed [{\color{blue}\onlinecite{Sirker1}}] that in some models, although the quench leads across a quantum phase transition but there are no Fisher zeroes. In another work, it is showed [{\color{blue}\onlinecite{Vajna1}}] that without crossing equilibrium critical lines, the Fisher zeroes exist.


Here, we considered the 1D spin-1/2 ITF model with added the Dzyaloshinskii-Moriya interaction. Using the fermionization technique, the Hamiltonian is exactly diagonalized. The DM interaction induces chiral phase in the ground state phase diagram of the model. Our consequences clearly show that the DMI while its presence  changes the energy spectrum of system but it can not have any effects in DQPT of the system and therefore, behavior of DQPT of the system remains as a ITF model.


\vspace{0.3cm}


\end{document}